\documentstyle[12pt]{article}
\title{CP violation in the two-Higgs-doublet model: an example}
\author{L.\ Lavoura\thanks{On leave of absence
from Universidade T\'ecnica de Lisboa,
Lisbon,
Portugal.
Address after 6 September 1994:
CFMC,
av.\ prof.\ Gama Pinto 2,
P-1699 Lisboa Codex,
Portugal} \\
\small Department of Physics, Carnegie-Mellon University, \\
\small Pittsburgh, Pennsylvania 15213, U.S.A.}
\begin{document}
\maketitle
\begin{abstract}
In a general two-scalar-doublet model without fermions,
there is a unique source of CP violation,
$ J_1 $,
in the gauge interactions of the scalars.
It arises in the mixing of the three neutral physical scalars $ X_1 $,
$ X_2 $ and $ X_3 $.
CP violation may be observed via different decay rates for
$ X_1 \rightarrow H^+ W^- $ and
$ X_1 \rightarrow H^- W^+ $
(or,
alternatively,
for $ H^+ \rightarrow X_1 W^+ $ and
$ H^- \rightarrow X_1 W^- $
--- depending on which decays are kinematically allowed).
I compute the part of those CP-violating decay-rate differences
which is proportional to $ J_1 $.
The CP-invariant final-state-interaction phase
is provided by the absorptive parts of the one-loop diagrams.
I check the gauge invariance of the whole calculation.
\end{abstract}

\vspace{5mm}

\section{Introduction}

There are general reasons for the interest in the possibility
of CP violation in the scalar sector.
CP violation is a necessary ingredient for the generation
of the baryon asymmetry of the Universe \cite{sakharov}.
It is believed that CP violation in the Kobayashi--Maskawa matrix
is not large enough to explain that asymmetry \cite{gavela}.
It has been speculated \cite{nelson}
that the scalar sector might provide the missing CP violation.

There have been studies of possible signatures
of CP violation in the scalar sector.
It has been remarked \cite{pomarol}
that the simultaneous presence of the three couplings $ Z^0 S_1 S_2 $,
$ Z^0 S_1 S_3 $,
and $ Z^0 S_2 S_3 $,
where $ S_1 $,
$ S_2 $ and $ S_3 $ are three neutral scalar fields in any model,
implies CP violation.
Similarly,
the simultaneous presence of the three couplings
$ S_1 Z^0 Z^0 $,
$ S_2 Z^0 Z^0 $,
and $ S_1 S_2 Z^0 $,
represents CP violation.
This is because the C quantum number of the $ Z^0 $ is $ - 1 $.
Another work \cite{nowakowski}
has considered various CP-violating Lagrangians including scalars,
fermions,
and vector bosons,
and has suggested looking for CP violation in the decay mode
$ S \rightarrow Z^0 W^+ W^- $
--- occurring when,
in the rest frame of the decaying neutral scalar $ S $,
the momentum distribution of the $ W^+ $ is not the same
as the momentum distribution of the $ W^- $ ---
or,
in a similar fashion,
in $ S \rightarrow Z^0 H^+ H^- $.
The first of these CP-violating asymmetries
has later been computed \cite{cvetic}
in the context of the two-Higgs-doublet model.
However,
the decay mode $ S \rightarrow Z^0 W^+ W^- $ is
phase-space disfavored as compared to the simpler decay modes
$ S \rightarrow W^+ W^- $ and $ S \rightarrow Z^0 Z^0 $.
Other studies \cite{soni} have concentrated on CP-violating phenomena
originating in the interplay of scalars and fermions,
in particular the effects of top-quark physics.

The aim of this work is the computation of a CP-violating asymmetry
in the two-Higgs-doublet model without any fermions.
The model has gauge symmetry SU(2)$\otimes$U(1),
which is spontaneously broken to the U(1) of electromagnetism
by the vacuum expectation values (VEV's) of the two Higgs doublets.
I look for CP violation involving solely the gauge interactions
of the scalars.
For simplicity,
I do not consider the presence of fermions,
which presence would lead to extra sources of CP violation,
both in the fermion sector,
and in the interplay of the fermion and the scalar sectors.
I also omit possible sources of CP violation
in the cubic and quartic interactions of the physical scalars.
Those scalars are two charged particles $ H^{\pm} $,
with mass $ m_H $,
and three neutral particles $ X_1 $,
$ X_2 $ and $ X_3 $,
with masses $ m_1 $,
$ m_2 $ and $ m_3 $,
respectively.
Besides,
the spectrum of the model includes the massive intermediate vector bosons
$ W^{\pm} $ and $ Z^0 $,
with masses $ m_W = 80 $ GeV and $ m_Z = 91 $ GeV respectively,
and the massless photon.
For a fairly large range of the masses of the scalars,
either the two decays $ X_1 \rightarrow H^+ W^- $
and $ X_1 \rightarrow H^- W^+ $,
or the two decays $ H^+ \rightarrow X_1 W^+ $ and
$ H^- \rightarrow X_1 W^- $,
are kinematically allowed
(the neutral scalars may be numbered so that $ X_1 $
is the scalar for which one of these couples of decays is allowed).
Then,
the possibility of a CP-violating difference between the rate of one decay
and the rate of its CP-conjugated decay exists.
It is my purpose to calculate that difference.

It has recently been observed \cite{silva}
that the two-Higgs-doublet model
has one and only one source of CP violation
in the gauge interactions of the scalars.
I describe it briefly.
Because the U(1) of electromagnetism is preserved in the symmetry breaking,
we can,
without loss of generality,
choose a basis for the two scalar doublets in which only one of them,
$ H_1 $,
has a VEV $ v $,
while the second one,
$ H_2 $,
does not have a VEV.
The two doublets in that basis can be written
\begin{equation}
H_1 = \left( \begin{array}{c}
G^+ \\ v + (H + i G^0) / \sqrt{2}
\end{array} \right)\, ,
\hspace{5mm}
H_2 = \left( \begin{array}{c}
H^+ \\ (R + i I) / \sqrt{2}
\end{array} \right)\, .
\label{doublets}
\end{equation}
$ G^+ $ and $ G^0 $ are the Goldstone bosons,
which become the longitudinal components of the $ W^+ $ and $ Z^0 $,
respectively.
$ H $,
$ R $,
and $ I $ are linear combinations
of the three neutral scalar fields $ X_1 $,
$ X_2 $,
and $ X_3 $,
which are the eigenstates of mass.
Those linear combinations are given by an orthogonal matrix $ T $,
\begin{equation}
\left( \begin{array}{c} H \\ R \\ I \end{array} \right) =
T\, \left( \begin{array}{c} X_1 \\ X_2 \\ X_3 \end{array} \right)\, .
\label{linearT}
\end{equation}
Without loss of generality,
we can assume $ T $ to have determinant $ +1 $.
Then,
the following useful identities follow:
\begin{equation}
T_{2i} T_{3j} - T_{2j} T_{3i}
= \sum_{k=1}^3 \epsilon_{ijk} T_{1k}\, ,
\label{epsilon}
\end{equation}
where $ \epsilon_{ijk} $ is the totally antisymmetric tensor
with $ \epsilon_{123} = +1 $.
There is CP violation in the gauge interactions of the scalars \cite{silva}
if and only if $ m_1 $,
$ m_2 $ and $ m_3 $ are all different,
and
\begin{equation}
J_1 \equiv T_{11} T_{12} T_{13}
\label{J1}
\end{equation}
is non-zero.
The quantity $ J_1 $ has in the two-Higgs-doublet model
a role analogous to the one of Jarlskog's \cite{jarlskog} $ J $
in the three-generation standard model.
Notice however that,
here,
there are in principle other sources of CP violation,
in the cubic and quartic interactions of the scalars.
I will neglect those extra sources of CP violation throughout this work.

It is important to remark that,
though $ J_1 $ represents CP violation in the mixing
of the three neutral scalars,
this source of CP violation has nothing to do with the fermions
and with the identification of,
say,
$ H $ and $ R $ as being scalars,
and $ I $ as being a pseudoscalar.
That identification can only be done when a specific Yukawa Lagrangian,
coupling the two scalar doublets to the fermion sector,
is considered,
which I do not do here.
Specifically,
as is clear from Eq.~\ref{J1},
$ R $ and $ I $ play a completely equivalent role in $ J_1 $
---indeed,
as long as there are no Yukawa couplings,
$ R $ and $ I $ may rotate into each other by a simple U(1) rephasing
of $ H_2 $.
Also,
$ J_1 $ cannot be the source of,
say,
CP violation in the kaon system.
If fermions are introduced in the model,
the mixing of the neutral scalars will in principle
lead to more CP violation than simply $ J_1 $,
because of the Yukawa interactions of the scalars with the fermions.

\section{General features of the calculation}

Let us consider how CP violation proportional to $ J_1 $
arises in the decay modes that I consider here.
All the vertices needed for the calculations in this paper
have been listed in Figure 1.
The tree-level diagram with incoming particles
$ W^- $,
$ H^+ $ and $ X_1 $
is proportional to $ i (T_{21} - i T_{31}) $,
while the diagram with incoming particles $ W^+ $,
$ H^- $ and $ X_1 $ is proportional to $ - i (T_{21} + i T_{31}) $.
Now take a look at Figure 2,
in which all the one-loop diagrams
which lead to CP violation when interfering with the tree-level diagram
are collected.
Consider for instance the first diagram,
with a loop of $ W^+ W^- $,
and then $ X_2 $,
as an intermediate state.
That diagram is equal to $ i^6 T_{11} T_{12} (T_{22} - i T_{32}) $,
times a certain momentum integral $ i I_k $.
The seven factors of $ i $ come,
three from the vertices,
three from the propagators,
and one from the Wick rotation in the momentum integral.
Therefore,
the interference of this diagram with the tree-level one
is proportional to the real part of $ (-i) (T_{21} + i T_{31})
i^6 T_{11} T_{12}
(T_{22} - i T_{32}) i I_k = T_{11} T_{12} (T_{11} T_{12}
+ i T_{13}) I_k $.
The $T$-matrix factor has an imaginary part equal to $ J_1 $.
Therefore,
if the momentum integral has an absorptive
({\it i.e.},
imaginary) part,
then the interference term will include $ J_1 $ times that absorptive part.
This is CP-violating.
The absorptive part of the integral plays in the calculation
the role of a CP-invariant final-state-interaction phase,
which allows $ J_1 $ to manifest itself.

We can check in a similar fashion that the absorptive parts of all other
nine one-loop diagrams in Figure 2 lead,
when one considers the interference of those diagrams with the tree-level one,
to CP violation.
Indeed,
a careful study of the model and all its vertices shows that the ten diagrams
in Figure 2 are the only ones which lead to CP violation
proportional to $ J_1 $ in this process.\footnote{There are other diagrams
which may also lead to CP violation in this process,
but which include other sources of CP violation,
in the cubic scalar interactions.
I neglect those extra sources of CP violation,
just as I neglect fermionic sources of CP violation.}
The CP violation manifests itself in a difference of the decay rates
of $ X_1 \rightarrow W^+ H^- $ and $ X_1 \rightarrow W^- H^+ $,
or of $ H^+ \rightarrow X_1 W^+ $ and $ H^- \rightarrow X_1 W^- $,
whichever pair of decays is kinematically allowed.

Let me be more explicit.
At tree-level,
the amplitude for the decay $ X_1 \rightarrow W^+ H^-$,
or for the decay $ H^+ \rightarrow X_1 W^+ $,
is $ (\epsilon_{\nu} P_H^{\nu}) i g (T_{21} - i T_{31}) $,
from the tree-level vertex.
Here,
$ \epsilon_{\nu} $ is the polarization vector of the outgoing $ W^+ $,
and $ P_H^{\nu} $ is the incoming momentum of the $ H^{\pm} $.
This is because the polarization vector of the $ W^+ $
is orthogonal to the momentum of that vector boson.
At one-loop level,
for the same reason,
each diagram in Figure 2 contributes
$ M_k = (\epsilon_{\nu} P_H^{\nu}) g^3 C_k i I_k $.
Here,
$ C_k $ are the various $ i $ factors and $T$-matrix factors
from the vertices and propagators in the diagram,
and $ i I_k $ is the momentum integral,
with the $ i $ coming from the Wick rotation.
The amplitudes for the CP-conjugated decays are,
at tree-level,
$ (\epsilon_{\nu} P_H^{\nu}) (-i) g (T_{21} + i T_{31}) $,
and,
at one-loop level from Figure 2,
$ (\epsilon_{\nu} P_H^{\nu}) g^3 (- C_k^{\ast}) i I_k $.
Notice that,
while the momentum integral is the same,
the vertex factors are complex-conjugated.
Then,
the CP-violating asymmetries are
\begin{eqnarray}
\frac{{\rm BR} (X_1 \rightarrow W^+ H^-)
- {\rm BR} (X_1 \rightarrow W^- H^+)}{{\rm BR} (X_1 \rightarrow W^+ H^-)
+ {\rm BR} (X_1 \rightarrow W^- H^+)}
& &
\nonumber\\*[1mm]
\left[
{\rm or}
\hspace{1mm}
\frac{{\rm BR} (H^+ \rightarrow X_1 W^+)
- {\rm BR} (H^- \rightarrow X_1 W^-)}{{\rm BR} (H^+ \rightarrow X_1 W^+)
+ {\rm BR} (H^- \rightarrow X_1 W^-)}
\right]
& \approx &
2 g^2\, \sum_{k=1}^{10}\,
\frac{{\rm Im} [(T_{21} + i T_{31}) C_k]\, {\rm Re} (i I_k)}{1 -
T_{11}^2}\, ,
\label{asymmetry}
\end{eqnarray}
where I used the orthogonality of $ T $ to write $ T_{21}^2 + T_{31}^2
= 1 - T_{11}^2 $.
The imaginary part of $ (T_{21} + i T_{31}) C_k $
is simply $ J_1 $ times a number
--- typically $ \pm 1 $ or $ \pm 2 $.
The momentum integral $ i I_k $ (the $ i $ is from the Wick rotation)
has a real part if cuts in the corresponding diagram
lead to absorptive parts.
Notice that in Eq.~\ref{asymmetry}
I have used the approximation of taking,
in the denominator,
only the square of the modulus of the tree-level contribution
to the amplitude.

It is clear from Eq.~\ref{asymmetry} that the asymmetry
will be of the form
\begin{equation}
g^2\, \frac{2 T_{11} T_{12} T_{13}}{1 - T_{11}^2}\, A\, ,
\label{form}
\end{equation}
where $ A $ represents the sum of the absorptive parts of all the diagrams
in Figure 2,
weighted by appropriate numbers $ \pm 1 $ or $ \pm 2 $
(see the preceding paragraph).
As one is interested in how large the asymmetry can be,
I now consider the mixing-matrix factor in Eq.~\ref{form}.
Because one is constrained by the orthogonality condition
$ T_{12}^2 + T_{13}^2 = 1 - T_{11}^2 $,
it is obvious that that factor will be maximal when $ |T_{12}| = |T_{13}| $,
and we will then have
\begin{equation}
\frac{2 T_{11} T_{12} T_{13}}{1 - T_{11}^2} = T_{11}\, .
\label{maximum}
\end{equation}
Clearly,
the order of magnitude of this quantity $ T_{11} $ is 1.
It is at this point important to remark that,
in this specific example,
the CP-violating asymmetry approaches its maximum
when the decay rate decreases.
Indeed,
as $ T_{11} \rightarrow 1 $,
the asymmetry becomes potentially larger
(as long as $ |T_{21}| $ remains equal to $ |T_{31}| $),
but the decay rate,
which is proportional to $ 1 - T_{11}^2 $,
approaches zero.
Similarly,
the decay rate becomes larger if $ T_{11} \rightarrow 0 $,
but then $ J_1 \rightarrow 0 $ and the CP asymmetry also vanishes.
This situation is reminiscent
of the case of CP-violating asymmetries in decay modes of the $ B^0 $ mesons,
which are generally predicted to be larger
when the branching ratios are smaller,
and vice-versa.

\section{Gauge invariance}

A way to check that the diagrams in Figure 2 are all the relevant ones
is to check whether their computation yields a gauge-invariant result.
Because I just want to compute the absorptive parts of the diagrams,
which are finite,
it is sufficient to compute the diagrams in the unitary gauge,
in which no Goldstone bosons and no ghosts are present.
However,
in a general 't-Hooft gauge,
in each of the diagrams in Figure 2,
a $ G^{\pm} $ can be used instead of the $ W^{\pm} $,
or a $ G^0 $ can be used instead of the $ Z^0 $.
In the two diagrams which have a loop only of $ W^{\pm} $ or of $ Z^0 $
(diagrams 1 and 2),
ghost loops must also be considered in an 't-Hooft gauge.
Now,
in an 't-Hooft gauge,
the $ W^{\pm} $ propagator contains an extra piece
(relative to the unitary gauge)
in which the $ W^{\pm} $ has an unphysical squared mass $ W $.
Similarly,
the charged Goldstone bosons $ G^{\pm} $
and the charged ghosts $ c^{\pm} $ have unphysical squared mass $ W $
in an 't-Hooft gauge.
When $ W $ is not infinite as in the unitary gauge,
each diagram by itself contains a $W$-dependent absorptive part.
However,
all those unphysical absorptive parts must cancel out
when one considers the whole set of diagrams.
The same thing can be said about the propagators of the $ Z^0 $,
which has a piece with unphysical squared mass $ Z $,
and of the Goldstone boson $ G^0 $ and ghost $ c^0 $,
which have squared mass $ Z $.
(In principle,
$ Z \neq W $.)
The whole set of propagators is given in Figure 3.
The sum over all diagrams of all the absorptive parts
must be independent of both $ W $ and $ Z $.
I have checked that independence.

To be sure,
gauge-independence only applies to an observable quantity.
Thus,
gauge-independence in this case only occurs when
1) the three external particles are all on mass shell,
that is,
$ P_H^2 = m_H^2 $,
$ P_W^2 = m_W^2 $,
and $ P_1^2 = m_1^2 $,
where $ P_H $,
$ P_W $ and $ P_1 $ are the incoming momenta of the external $ H^{\pm} $,
$ W^{\pm} $,
and $ X_1 $,
respectively;
2) one suppresses from the amplitude all terms proportional to $ P_W^{\nu} $,
because the amplitude must be multiplied by the polarization vector
$ \epsilon_{\nu} $ of the $ W^{\pm} $,
and $ \epsilon_{\nu} P_W^{\nu} = 0 $;
3) one considers,
from each one-loop diagram,
only the part which is proportional to $ J_1 $ upon interference
with the tree-level amplitude;
4) one only considers the absorptive part of each one-loop diagram.
One has also to take into consideration that the gauge-dependent
absorptive parts sometimes cancel between two similar diagrams
with intermediate virtual particle $ X_3 $ instead of $ X_2 $,
whenever those absorptive parts do not depend on the mass of that
intermediate particle ($ m_3 $ or $ m_2 $);
while other absorptive parts cancel among different-looking diagrams.

\section{Contribution of each diagram}

I present in this section the results of the computation
of each of the ten diagrams in Figure 2.

I am only interested in the absorptive part of each diagram.
Those absorptive parts are simpler to compute
when the power of the integration momentum $ k $ in the denominator
is lower;
in particular,
if only one propagator $ k^2 - m^2 $ occurs in the denominator,
there is no absorptive part.
Moreover,
I want to display explicitly how the cancellations
which lead to gauge invariance of the whole result occur.
Therefore,
my method has been the following.
First I have added,
for each particular diagram,
the contributions with gauge bosons,
the ones with Goldstone bosons,
and the ones with ghosts.
Then,
I have simplified as much as possible the numerators of the integrands,
in such a way that factors in some terms of those numerators cancel out
some of the propagators in the denominators,
thereby reducing the overall power of the integration momentum $ k $
in the denominator.
As a result,
each diagram becomes the sum of three parts:
one piece with an overall $ k^2 $ in the denominator,
leading to no absorptive part;
one piece with $ k^4 $ in the denominator;
and --- for the diagrams 7 through 10 only ---
one piece with $ k^6 $ in the denominator.
I write down here the two latter pieces only,
because they are the only ones
which lead to absorptive parts in the integrals.

For each diagram,
I have taken separately the factors with $T$-matrix elements,
and all the $ i $ factors from both the vertices and the propagators,
multiplied that by the factor $ (T_{21} + i T_{31}) $
arising in the interference with the tree-level diagram,
and taken the imaginary part of the result.
That imaginary part is always a multiple of $ J_1 $.
This I did because it is only the quantities
$ {\rm Im} [(T_{21} + i T_{31}) C_k] $
which are relevant for the computation of the asymmetry,
according to Eq.~\ref{asymmetry}.
Also,
it is only for these quantities
$ {\rm Im} [(T_{21} + i T_{31}) C_k] $
that the cancellation of the gauge dependence must occur.
As a consequence,
each momentum integral presented below should be looked upon as being
an $ i I_k $ in the notation of section 2
(the $ i $ arising when the Wick rotation is performed),
and only the absorptive parts of each such momentum integral
are meaningful in this context.

Each diagram in Figure 2 is in reality two diagrams,
one in which an intermediate state $ X_2 $ appears,
and another one in which the intermediate state is $ X_3 $.
Those two diagrams have,
as is easily seen,
opposite signs of $ J_1 $ in the interference term
$ {\rm Im} [(T_{21} + i T_{31}) C_k] $.
Therefore,
if the corresponding momentum integrals happen not to depend
on the masses $ m_2 $ or $ m_3 $
--- because the propagator of $ X_2 $ or of $ X_3 $
might have been cancelled out by a similar factor
in the numerator of the integrand ---
then those terms become irrelevant for the present computation.
As a consequence,
for each diagram,
I only write down those integrals which do depend
on $ m_2 $ or $ m_3 $.

As before,
$ P_1 $,
$ P_W $ and $ P_H $ are the incoming momenta of the scalar $ X_1 $,
the gauge boson $ W^- $,
and the charged scalar $ H^+ $.
I have systematically set all three external particles on mass-shell,
and used whenever possible $ \epsilon_{\nu} P_W^{\nu} = 0 $
in order to simplify the amplitudes.

I first define the various combinations of the integration momentum $ k $
and of the incoming momenta,
and of the masses,
which appear in the denominators of the momentum integrals.
They are
\begin{eqnarray}
D_1 & \equiv & k^2 - m_W^2\, ,
\label{d1}\\
D_2 & \equiv & k^2 + 2 k \cdot P_1 + m_1^2 - m_W^2\, ,
\label{d2}\\
D_3 & \equiv & k^2 + 2 k \cdot P_1 + m_1^2 - m_H^2\, ,
\label{d3}\\
D_4 & \equiv & k^2 - 2 k \cdot P_H + m_H^2 - m_2^2\, ,
\label{d4}\\
D_5 & \equiv & k^2 + 2 k \cdot P_W + m_W^2 - m_Z^2\, ,
\label{d5}\\
D_6 & \equiv & k^2 + 2 k \cdot P_W - 2 k \cdot P_H - m_1^2 + 2 m_W^2 + m_H^2\,
{}.
\label{d6}
\end{eqnarray}
Notice that only $ D_4 $ depends on $ m_2 $.

It is convenient to define the ``triangular function'' $ \lambda $:
\begin{equation}
\lambda(A, B, C) \equiv A^2 + B^2 + C^2 - 2 (A B + A C + B C)\, .
\label{lambda}
\end{equation}
This function is negative if and only if one can form a triangle
with sides of length $ \sqrt{A} $,
$ \sqrt{B} $,
and $ \sqrt{C} $.

I now present the results for the first three diagrams.
Diagram 1:
\begin{eqnarray}
M_{1} & = &
\epsilon_{\nu} P_H^{\nu}\, g^3\, J_1\,
\frac{m_3^2 - m_2^2}{(m_1^2 - m_2^2)(m_1^2 - m_3^2)}\,
\frac{- m_1^4 + 4 m_1^2 m_W^2 - 12 m_W^4}{4 m_W^2}
\int \frac{d^4 k}{(2 \pi)^4}\, \frac{1}{D_1 D_2}\, .
\label{m1}
\end{eqnarray}
Diagram 2:
\begin{eqnarray}
M_{2} & = &
\epsilon_{\nu} P_H^{\nu}\, g^3\, J_1\,
\frac{m_3^2 - m_2^2}{(m_1^2 - m_2^2)(m_1^2 - m_3^2)}
\nonumber\\
      & \times &
\frac{- m_1^4 + 4 m_1^2 m_Z^2 - 12 m_Z^4}{8 m_Z^2}
\int \frac{d^4 k}{(2 \pi)^4}\, \frac{1}{D_1 (m_W \rightarrow m_Z)
D_2 (m_W \rightarrow m_Z)}\, .
\label{m2}
\end{eqnarray}
(There is an extra factor $ 1/2 $ in diagram 2 relative to diagram 1,
which is a symmetry factor,
due to the two identical $Z^0$'s in the loop.)
Diagram 3:
\begin{eqnarray}
M_{3} & = &
\epsilon_{\nu} P_H^{\nu}\, g^3\, J_1\,
\frac{m_3^2 - m_2^2}{(m_1^2 - m_2^2)(m_1^2 - m_3^2)}\,
\frac{\lambda(m_1^2, m_W^2, m_H^2)}{2 m_W^2}\,
\int \frac{d^4 k}{(2 \pi)^4}\,
\frac{1}{D_1 D_3}\, .
\label{m3}
\end{eqnarray}
The results for each of these three diagrams
are separately gauge-invariant.
This is because the gauge-dependent piece in each of them
(the piece depending on either $ W $ or $ Z $) is the same,
for each case,
in the diagram with an intermediate $ X_2 $,
and in the diagram with an intermediate $ X_3 $.
As the sign of the $ J_1 $ factor is opposite,
the gauge-dependent piece cancels out between the diagrams with intermediate
$ X_2 $ and those with intermediate $ X_3 $.
The same phenomenon partially occurs in all other diagrams but,
in each of them individually,
some gauge dependence always remains,
as is seen in the following.

Diagram 4:
\begin{eqnarray}
M_{4} & = &
- \epsilon_{\nu}\, g^3\, J_1\,
\int \frac{d^4 k}{(2 \pi)^4}\,
\left\{
\frac{P_H^{\nu} (m_H^2 - m_2^2)}{4 m_W^2 D_4 (k^2 - W)}
+ \frac{k^{\nu} (m_W^2 + m_H^2 - m_2^2)}{2 D_1 D_4 (m_H^2 - m_W^2)}
\right.
\nonumber\\
      & - &
\left.
P_H^{\nu}\, \frac{\lambda(m_2^2, m_W^2, m_H^2) + 4 m_W^4}{4 D_1 D_4 m_W^2
(m_H^2 - m_W^2)}
- (m_2 \rightarrow m_3)
\right\}\, .
\label{m4}
\end{eqnarray}
Diagram 5:
\begin{eqnarray}
M_{5} & = &
- \epsilon_{\nu}\, g^3\, J_1\,
\frac{m_Z^2 - m_W^2}{m_W^2}\,
\int \frac{d^4 k}{(2 \pi)^4}\,
\left\{ k^{\nu}\, \frac{m_2^2 - m_1^2}{4 m_Z^2 D_4 [(k + P_W)^2 - Z]}
\right.
\nonumber\\
      & + &
\left. \frac{1}{D_4 D_5}\,
\left(
- \frac{P_H^{\nu}}{2}\, + k^{\nu}\, \frac{m_Z^2 - m_2^2 + m_1^2}{4 m_Z^2}
\right) - (m_2 \rightarrow m_3) \right\}\, .
\label{m5}
\end{eqnarray}
Diagram 6:
\begin{eqnarray}
M_{6} & = &
\epsilon_{\nu} P_H^{\nu}\, g^3\, J_1\, \frac{1}{4}\,
\int \frac{d^4 k}{(2 \pi)^4}\,
\left\{
\frac{m_2^2 - m_1^2}{m_W^2}\,
\frac{1}{D_4 [(k + P_W)^2 - Z]}
\right.
\nonumber\\
      & + &
\left.
\frac{\lambda(m_Z^2, m_1^2, m_2^2)}{D_4 D_5 m_W^2 (m_1^2 - m_3^2)}\,
- (m_2 \leftrightarrow m_3)
\right\}\, .
\label{m6}
\end{eqnarray}
Diagram 7:
\begin{eqnarray}
M_{7} & = &
- \epsilon_{\nu}\, g^3\, J_1\, \frac{2 m_W^2 - m_Z^2}{4 m_W^2}\,
\int \frac{d^4 k}{(2 \pi)^4}\, (P_H - k)^{\nu}\,
\left\{
\frac{(m_1^2 - m_2^2)}{m_Z^2 D_4 [(k + P_W)^2 - Z]}
\right.
\nonumber\\
      & + &
\left.
\frac{1}{D_4 D_6}
+ \frac{m_2^2 - m_1^2 - m_Z^2}{m_Z^2}\, \frac{1}{D_4 D_5}
+ (2 m_W^2 + m_Z^2 - 2 m_H^2 - m_1^2 - m_2^2)
\frac{1}{D_4 D_5 D_6}
\right.
\nonumber\\
      & - &
\left.
(m_2 \leftrightarrow m_3)
\right\}\, .
\label{m7}
\end{eqnarray}
Diagram 8:
\begin{eqnarray}
M_{8} & = &
\epsilon_{\nu}\, g^3\, J_1\, \frac{1}{2 m_W^2}\,
\int \frac{d^4 k}{(2 \pi)^4}\, (k - P_H)^{\nu}\,
\left[
\frac{(m_H^2 - m_2^2)}{D_4 (k^2 - W)}
\right.
\nonumber\\
      & + &
\left.
\frac{m_2^2 - m_H^2 - m_W^2}{D_1 D_4}
+ \frac{m_W^2}{D_3 D_4}
+ \frac{1}{D_1 D_3 D_4}\,
(3 m_W^4 - m_H^4 + m_1^2 m_H^2 - m_1^2 m_2^2
\right.
\nonumber\\
      &   &
\left.
+ m_H^2 m_2^2 - m_1^2 m_W^2 - m_2^2 m_W^2 - 2 m_W^2 m_H^2)\,
- (m_2 \leftrightarrow m_3)
\right]\, .
\label{m8}
\end{eqnarray}
Diagram 9:
\begin{eqnarray}
M_{9} & = &
- \epsilon_{\nu}\, g^3\, J_1\,
\int \frac{d^4 k}{(2 \pi)^4}\,
\left\{
\frac{(k - P_H)^{\nu}}{4 m_W^2}\,
\frac{(m_H^2 - m_2^2)}{D_4 [(k + P_1)^2 - W]}\,
+ \frac{(k - P_H)^{\nu}}{4 m_W^2}\,
\frac{(m_H^2 - m_2^2)}{D_4 (k^2 - W)}
\right.
\nonumber\\
      & + &
\frac{(k - P_H)^{\nu}}{4}\,
\left( \frac{1}{D_2 D_4}\, - \frac{1}{D_1 D_4} \right)
- \frac{(k - P_H)^{\nu}}{4}\, \frac{m_H^2 - m_2^2}{m_W^2}\,
\left( \frac{1}{D_2 D_4}\, + \frac{1}{D_1 D_4} \right)
\nonumber\\
      & + &
\frac{1}{D_1 D_2 D_4}\,
\left[
\frac{- 6 m_W^2 + 2 m_H^2 + m_1^2}{4}\, P_H^{\nu}
+ \frac{4 m_W^2 - 2 m_2^2 - m_1^2}{4}\, k^{\nu}
\right.
\nonumber\\
      & + &
\left.
\left.
\frac{m_1^2 (m_H^2 - m_2^2)}{4 m_W^2}\, (k - P_H)^{\nu}
\right]
- (m_2 \leftrightarrow m_3)
\right\}\, .
\label{m9}
\end{eqnarray}
Diagram 10:
\begin{eqnarray}
M_{10} & = &
- \epsilon_{\nu}\, g^3\, J_1\, \frac{1}{4}\,
\int \frac{d^4 k}{(2 \pi)^4}
\left\{
\frac{1}{(k^2 - W) [(k + P_W)^2 - Z]}\, k^{\nu}\, \frac{m_2^2}{m_W^2}
\right.
\nonumber\\
      & + &
\frac{1}{D_4 [(k + P_W)^2 - Z]}\, (k - 2 P_H)^{\nu}\,
\frac{m_1^2 - m_2^2}{m_Z^2}\,
+ \frac{1}{D_4 (k^2 - W)}\, (k - 2 P_H)^{\nu}\,
\frac{m_H^2 - m_2^2}{m_W^2}
\nonumber\\
      & - &
k^{\nu}\,
\left(
\frac{1}{D_1 D_4}\, + \frac{1}{D_4 D_5}
\right)
+ k^{\nu}\, \frac{1}{D_1 D_4 D_5}\,
(4 m_2^2 - 2 m_W^2 + 2 m_1^2 + 2 m_H^2 - m_Z^2)\,
\nonumber\\
      & + &
k^{\nu}\, \frac{(m_2^2 - m_H^2)(m_1^2 - m_2^2)}{m_W^2}\,
\frac{1}{D_1 D_4 D_5}\,
- k^{\nu}\, \frac{m_2^2}{m_W^2 D_1 D_5}
\nonumber\\
      & + &
(2 P_H - k)^{\nu}\,
\left[
\frac{3 m_2^2 + 2 m_W^2 - m_1^2 - 2 m_H^2}{D_1 D_4 D_5}\,
+ \frac{m_1^2 - m_2^2}{m_Z^2}\, \frac{1}{D_4 D_5}
\right.
\nonumber\\
      & + &
\left.
\left.
\frac{m_H^2 - m_2^2}{m_W^2}\,
\left(
\frac{1}{D_1 D_4}\, + \frac{m_Z^2}{D_1 D_4 D_5}
\right)
\right]\,
- (m_2 \rightarrow m_3)
\right\}\, .
\label{m10}
\end{eqnarray}

It is simple to check now that most
terms depending on either of the unphysical squared-masses
$ W $ or $ Z $ cancel among $ M_4 $ through $ M_{10} $.
The only two exceptions are the first terms
in the curly brackets in the expressions for $ M_9 $
and for $ M_{10} $.
The term in $ M_9 $ contains the integral of
$ (k - P_H)^{\nu} / D_4 [(k + P_1)^2 - W] $,
while the one in $ M_{10} $ contains the integral of
$ k^{\nu} / (k^2 - W) [(k + P_W)^2 - Z] $.
However,
it is easily found that the absorptive parts of
both these integrals are proportional to $ P_W^{\nu} $,
and therefore give a vanishing contribution to the amplitude.
This ensures the gauge invariance of the whole computation.
In each of Eqs.~\ref{m4} to \ref{m10},
therefore,
one only has to consider only the terms independent of $ W $ and of $ Z $.
Those terms are the ones that would have been obtained
had the computation been performed directly in the unitary gauge.
Some of those terms,
however,
still yield zero absorptive contributions.

\section{Absorptive parts}

In this section I briefly review
the computation of the absorptive parts of the integrals.

Consider first an integral with $ k^4 $ in the denominator:
\begin{equation}
i I_4 = \int \frac{d^4 k}{(2 \pi)^4}
\frac{r k^{\nu} + s P_H^{\nu}}{[(k + P_X)^2 - m_A^2][(k + P_Y)^2
- m_B^2]}\, ,
\label{I4}
\end{equation}
where $ r $ and $ s $ are some coefficients.
Let $ (P_X - P_Y)^2 \equiv m_C^2 > 0 $.
The integral is divergent.
After introducing a Feynman parameter $ x $,
and using dimensional regularization,
one obtains
\begin{equation}
i I_4 = \frac{i}{16 \pi^2}\, \int_0^1 dx
\left[
r (P_Y - P_X)^{\nu} x - r P_Y^{\nu} + s P_H^{\nu}
\right]\,
\left(
\frac{2}{4-d}\, - \gamma - \ln \frac{\Delta}{4 \pi \mu^2}
\right)\, ,
\label{result4}
\end{equation}
where
\begin{equation}
\Delta = m_C^2 x^2 + (m_A^2 - m_B^2 - m_C^2) x + m_B^2\, .
\label{Delta4}
\end{equation}
The absorptive part arises when $ \Delta $ is negative for
part of the integration domain of $ x $.
This happens when $ m_C $ is larger than $ m_A + m_B $.
Substituting $ \ln ( \Delta / 4 \pi \mu^2 ) $ by $ - i \pi $
and integrating over the part of the interval $ [0, 1] $
in which $ \Delta $ is negative,
one gets
\begin{eqnarray}
{\rm Re}\, (i I_4) & = & - \frac{1}{16 \pi}\,
\frac{\sqrt{\lambda ( m_A^2, m_B^2, m_C^2 )}}{m_C^2}
\nonumber\\
               & \times &
\left\{
r
\left[
- \frac{(P_X + P_Y)^{\nu}}{2}
+ \frac{(P_X - P_Y)^{\nu}}{2} \frac{m_A^2 - m_B^2}{m_C^2}
\right]
+ s P_H^{\nu}
\right\}\, .
\label{absorptive4}
\end{eqnarray}

Using this result,
one sees that the absorptive parts of some of the integrals
in the previous section are proportional to $ P_W^{\nu} $,
and therefore irrelevant.
This is true not only of the two gauge-dependent integrals
which did not cancel out,
as I pointed out in the last paragraph of the previous section;
it is also true of the absorptive parts of the integrals of
$ k^{\nu} / (D_1 D_5) $ (which arises in $ M_{10} $),
of $ (k - P_H)^{\nu} / (D_2 D_4) $ (in $ M_9 $),
of $ (k - P_H)^{\nu} / (D_3 D_4) $ (in $ M_8 $),
and of $ (k - P_H)^{\nu} / (D_4 D_6) $ (in $ M_7 $).
Eliminating all of these,
it is seen that the only remaining integrals of the type $ I_4 $
either have denominator $ D_1 D_4 $,
or have denominator $ D_4 D_5 $.
The first ones have an absorptive part if $ m_H > m_W + m_2 $,
and therefore $ H^{\pm} $ can decay to $ W^{\pm} X_2 $,
and the second ones have an absorptive part if $ m_1 > m_Z + m_2 $,
{\it i.e.},
if $ X_1 \rightarrow Z^0 X_2 $ is possible.

Now consider the integrals with $ k^6 $ in the denominator.
In our problem,
they are all of the form
\begin{equation}
i I_6 = \int \frac{d^4 k}{(2 \pi)^4}
\frac{\epsilon_{\nu} (r k^{\nu} + s P_H^{\nu})}{(k^2 - m_A^2)
[(k + P_1)^2 - m_B^2] [(k - P_H)^2 - m_C^2]}\, ,
\label{I6}
\end{equation}
where $ r $ and $ s $ are coefficients.
Specifically,
for the integral of $ 1 / (D_4 D_5 D_6) $ present in diagram 7
we have $ m_A = m_Z $,
$ m_B = m_2 $ and $ m_C = m_H $;
for the integral of $ 1 / (D_1 D_3 D_4) $ present in diagram 8
we have $ m_A = m_W $,
$ m_B = m_H $ and $ m_C = m_2 $;
for the integral of $ 1 / (D_1 D_2 D_4) $ present in diagram 9
it is $ m_A = m_B = m_W $,
and $ m_C = m_2 $;
and,
for the integral of $ 1 / (D_1 D_4 D_5) $ present in diagram 10,
it is $ m_A = m_2 $,
$ m_B = m_Z $ and $ m_C = m_W $.
The integral $ I_6 $ is finite and,
eliminating its part proportional to $ P_W^{\nu} $,
one obtains
\begin{equation}
i I_6 = \frac{i}{16 \pi^2}\, \epsilon_{\nu} P_H^{\nu}
\int_0^1 dx\, \int_0^1 dy\, (r y^2 + s y)
\frac{-1}{\Delta - i \epsilon^+}\, ,
\label{result6}
\end{equation}
where
\begin{eqnarray}
\Delta & = & A_Y y^2 + B_Y y + m_A^2\, ,
\label{Delta6}\\
A_Y & \equiv & m_W^2 x^2 + (m_1^2 - m_H^2 - m_W^2) x + m_H^2\, ,
\label{eq:AY}\\
B_Y & \equiv & (m_H^2 - m_1^2 + m_B^2 - m_C^2) x + m_C^2 - m_A^2 - m_H^2\, ,
\label{eq:BY}
\end{eqnarray}
and $ \epsilon^+ $ in a positive infinitesimal quantity.
Using now the fact that the imaginary part of $ 1 / (\Delta - i \epsilon^+) $
is $ i \pi \delta (\Delta) $,
and then integrating over the Feynman parameter $ y $,
thereby getting rid of the Dirac $ \delta $ function,
one gets
\begin{equation}
{\rm Re}\, (i I_6) = \frac{\epsilon_{\nu} P_H^{\nu}}{16 \pi}\,
\int dx
\left(
s \frac{- B_Y}{A_Y}
+ r \frac{B_Y^2}{A_Y^2}
- 2 r \frac{m_A^2}{A_Y}
\right)\,
\frac{1}{\sqrt{(B_Y)^2 - 4 m_A^2 A_Y}}\, .
\label{absorptive6}
\end{equation}
Here,
the integral must be performed over that part of the interval
$ [0, 1] $ for which the two values of $ y $
defined by the condition $ \Delta = 0 $ are positive
and smaller than 1.
That sub-interval of $ [0, 1] $
must be found for each of the four integrals of the type $ I_6 $
individually.
It is larger or smaller according to whether
some cuts can be made in the corresponding Feynman diagram:
namely,
whether $ m_H > m_A + m_C $ or not,
and whether $ m_1 > m_A + m_B $ or not.
In any case,
the integral over $ x $ can be performed analytically,
and it yields a rather complicated logarithmic function.

\section{Results}

We now have all the ingredients needed to compute the asymmetry.
The asymmetry is equal to $ g^2 \approx 0.43 $ times
a mixing-matrix factor,
studied in section 2,
which should be of order $ 0.1 $ to $ 1 $,
times the sum $ A $ of all absorptive parts,
which itself includes a suppression factor $ 1 / (16 \pi) $.

Now,
one should note that the absorptive parts of diagrams 1,
2,
3 and 6 all diverge when $ m_2 $ (or $ m_3 $) approach $ m_1 $.
This is simply because in those diagrams one has a scalar $ X_2 $
(or $ X_3 $) propagating with momentum $ P_1 $ such that $ P_1^2 = m_1^2 $.
Those divergences do not cancel in the absorptive parts,
because the specific values of those absorptive parts depend
on different parameters:
in diagram 1,
on the decay width $ X_1 \rightarrow W^+ W^- $,
in diagram 2,
on the decay width $ X_1 \rightarrow Z^0 Z^0 $,
in diagram 3,
on the decay width $ X_1 \rightarrow H^+ W^- $,
and,
in diagram 6,
on the decay width $ X_1 \rightarrow Z^0 X_2 $ or $ Z^0 X_3 $.
Of course,
we know that these divergences are not genuine,
they might be eliminated by a proper treatment in which one would take into
account the finite width of the propagating $ X_2 $ or $ X_3 $,
and besides,
from a different line of reasoning \cite{silva},
one knows anyway that CP violation disappears
and $ J_1 $ loses its meaning
when the masses of any two of the three neutral scalars become equal.
A proper treatment of these divergences
at $ m_1 = m_2 $ or $ m_1 = m_3 $ would lead me far astray,
and therefore I simply avoided,
in general,
considering the region of the parameter space in which either
$ m_2 $ or $ m_3 $ are close to $ m_1 $.

Avoiding those regions in which the present approximation loses its
validity,
I find that the sum $ A $ of all absorptive parts is typically
of order of magnitude $ 10^{-3} $ or $ 10^{-2} $.
A few examples are presented in the following table.

\begin{tabular}{||c|c|c|c||c||}
\hline
$m_H$ (GeV) & $m_1$ (GeV) & $m_2$ (GeV) & $m_3$ (GeV) & A \\
\hline
300 & 150 & 250 & 60 & $ - 5.23 \times 10^{-2} $ \\
200 & 100 & 70 & 450 & $ 1.08 \times 10^{-2} $ \\
200 & 500 & 100 & 150 & $ - 1.90 \times 10^{-2} $ \\
250 & 500 & 250 & 80 & $ 5.61 \times 10^{-2} $ \\
300 & 100 & 200 & 400 & $ 5.28 \times 10^{-3} $ \\
\hline
\end{tabular}

It is worth remarking that the total absorptive part $ A $
is always the final result of substantial cancellations
among the absorptive parts,
with different signs,
of the various individual diagrams.

\section{Conclusions}

In this paper I have presented a model calculation of a CP-violating
asymmetry in the two-scalar-doublet model.
The asymmetry chosen has been the different decay rates for
$ X_1 \rightarrow H^+ W^- $ and for
$ X_1 \rightarrow H^- W^+ $.
I believe this to be a quite interesting place
to look for CP violation in the scalar sector,
even if the present calculation turns out not to be very relevant.
This might happen mainly because
I have taken into account only one source of CP violation,
$ J_1 $ in the gauge interactions of the scalars,
while I neglected further sources of CP violation,
in the cubic scalar interactions
and in the Yukawa interactions with the fermions.
Even within the context of the approximation that I have used,
in which the decay proceeds essentially because of a tree-level vertex,
while the CP violation arises because of the interference
of that tree-level vertex
with one-loop diagrams with absorptive parts,
those further sources of CP violation will in principle lead
to extra contributions to the total asymmetry.

My interest here has been to ilustrate the specific way in which
$ J_1 $ arises in the computation of a CP asymmetry.
I observed that there is a kind of balance between the CP asymmetry
and the decay rate in this specific case
--- but only if the only source of CP violation is taken to be $ J_1 $ --,
with a large asymmetry being possible only when the decay rate is small,
and vice-versa.

I found that the asymmetry can attain values of order $ 10^{-2} $.
These values would increase or decrease if the interference with other
sources of CP violation in this mode were constructive or destructive.

Because of the presence of gauge bosons
in the internal lines of the one-loop diagrams that I had to compute,
I found it convenient to check the gauge invariance of the whole calculation.
I checked that the fictitious masses that appear in the propagators
of the $ W^{\pm} $ and of the $ Z^0 $,
and of the corresponding Goldstone bosons and ghosts,
in a general 't Hooft gauge,
lead to gauge-dependent absorptive parts for the individual diagrams,
which however cancel out when all the diagrams
which lead to CP violation proportional to $ J_1 $ are considered.
This constitutes a good check that one did not omit any diagram.

\vspace{2mm}

I thank Lincoln Wolfenstein
for suggesting to me
the computation in this work,
and for reading the manuscript.
Ling-Fong Li and Martin Savage
made useful technical suggestions.
This work was initiated during visits
to the Stanford Linear Accelerator Center,
and to the Institute for Nuclear Theory at the University of Washington,
the hospitality of which I acknowledge.
This work was supported by the United States Department of Energy,
under the contract DE-FG02-91ER-40682.

\vspace{10mm}

%

\vspace{10mm}

\hspace{5mm} {\bf FIGURE CAPTIONS}

\vspace{5mm}

Figure 1:
The vertices needed for the computation of the diagrams in Figure 2,
and the Feynman rules for their values.
All the particles are incoming particles,
and all the momenta are incoming momenta.
The indices $ k $,
$ j $ and $ l $ may be $ 1 $,
$ 2 $ or $ 3 $.
The meaning of the parameters $ W $ and $ Z $ is explained
in the caption of figure 3.
Other vertices which might at first sight be relevant,
like $ W^- W^+ G^0 $ and $ H^- H^+ G^0 $,
do not exist.

\vspace{5mm}

Figure 2:
The one-loop diagrams with external incoming particles $ W^- $,
$ H^+ $ and $ X_1 $,
which lead to CP violation upon interfering with the tree-level diagram,
if the integrals have absorptive parts.
In each case,
the $ W^{\pm} $ and the $ Z^0 $ in internal lines may be substituted
by the corresponding Goldstone bosons $ G^{\pm} $ and $ G^0 $,
respectively;
in the first two diagrams,
they may be substituted by ghost loops as well.

\vspace{5mm}

Figure 3:
The propagators relevant for the computation of the diagrams in Figure 2.
$ W $ and $ Z $ are unphysical parameters with dimension of squared mass,
which also arise in the ghost vertices in Figure 1.
The final physical results are independent of both $ W $ and $ Z $.

\end{document}